\documentclass[vecphys]{svmult}

\usepackage{makeidx}         
\usepackage{graphicx}        
\usepackage{multicol}        

\makeindex             

\begin{document}

\title*{Recent Advances in Studies of Current Noise}
\author{Ya.~M.~Blanter}
\institute{Kavli Institute of Nanoscience, Delft University of
 Technology, Lorentzweg 1, 2628 CJ Delft, The Netherlands\\
\texttt{blanter@tnntw14.tn.tudelft.nl}}

\maketitle

\abstract{This is a brief review of recent activities in the
field of current noise intended for newcomers. We first briefly
discuss main properties of shot noise in nanostructures, and then turn
to recent developments, concentrating on issues related to
experimental progress: non-symmetrized cumulants and quantum noise;
counting statistics; super-Poissonian noise; current noise and
interferometry.} 

\section{Introduction}
\label{sec:intro}

Current noise in the last decade proved to be an efficient means
of investigation of nanostructures. Currently, it is a broad
field, with over a hundred groups, experimental as well as
theoretical, and is actively developing. This short review
presents a brief introduction to the field, concentrating on
recent developments. There is a large amount of literature
available. General introduction to noise in solids can be found in
the broad scope book by Kogan \cite{Kogan}. An extensive review on
shot noise was written by B\"uttiker and the author \cite{BB00}. A
collection of shorter review articles intended to summarize the
main directions of the field was published as proceedings of the
NATO ARW held in Delft in 2002 \cite{Nazarov}. These publications
cover the field comprehensively, and there is no need to repeat
all the material. This article is intended for researchers wishing
to enter the field. We will only give a brief introduction to the
subject of shot noise, turning then to recent developments of the
field. The size of this article makes it impossible to describe
all research. We will therefore self-impose the following
limitations. First, we only consider papers published in 2000 and
later --- everything before that date can be found in Ref.
\cite{BB00}. Second, the choice of topics is mainly related to the
experimental breakthroughs. Many papers of excellent quality will
have to stay outside the framework of this article. Additionally,
on purpose we do not discuss here very important issues of
entanglement and the theory of measurement: In our opinion, they
are better discussed in connection with properties of qubits, and
we do not have enough space here for a comprehensive review of the
field.

Now we give a very brief overview of results well established in
the field. Current through any nanostructure fluctuates in time.
There are at least two reasons for these fluctuations: (i) thermal
fluctuations of occupation numbers in the reservoirs; (ii)
randomness of transmission and reflection of electrons. At
equilibrium, only the former are important, and one has {\em
Nyquist-Johnson noise}. We define the noise spectral power,
\begin{equation} \label{noisedef}
S(\omega) = \left\langle \delta \hat{I} (t) \delta \hat{I} (t') +
\delta \hat{I} (t') \delta\hat{I} (t) \right\rangle_{\omega},
\end{equation}
where $\delta \hat{I} (t) \equiv \hat{I} (t) - \langle I \rangle$,
$\hat{I}$ is the current operator, and the averaging is both
quantum-mechanical and statistical over the states in the
reservoirs. Nyquist noise $S(0) = 4Gk_BT$, with $G$ being the
conductance of a nanostructure, just follows from the
fluctuation-dissipation theorem.

At zero temperature, only fluctuations due to the randomness of
scattering are important. They are known as {\em shot noise} and
can be expressed \cite{Khlus,Lesovik1,Buttiker90,MartinLandauer} in
terms of the transmission
eigenvalues $\{ T_p \}$ of the nanostructure, where $p$ label the
transport channels,
\begin{equation} \label{shotres}
S(0) = \frac{2_s e^3 \vert V \vert}{\pi\hbar} \sum_p T_p (1 -
T_p),
\end{equation}
and $2_s$ is the number of spin projections. Fully open ($T_p=1$)
and fully closed ($T_p = 0$) channels do not produce any noise,
since scattering is not random: electrons are either fully
reflected or fully transmitted.

To appreciate Eq. (\ref{shotres}), we need a reference point. The
latter was provided as early as 1918 by Schottky. Consider the
{\em Poisson process}: electrons enter a reservoir in a random and
uncorrelated fashion. In other words, the current is expressed as
$I(t) = e\sum \delta(t - t_n)$, where $t_n$ are random
uncorrelated quantities, with the average interval $\tau$ between
arrivals of consequent electrons. The average current is $I =
e/\tau$, while the current noise for this process is $S(\omega) =
2eI$ and does not depend on frequency. This {\em Poisson value}
$S_P = 2eI$ gives us the reference point. Taking into account
Landauer formula for conductance,
\begin{displaymath}
G = \frac{2_se^2}{2\pi\hbar} \sum_p T_p \ ,
\end{displaymath}
we see that the actual noise (\ref{shotres}) is always suppressed
with respect to $S_P$. This suppression is characterized by the
{\em Fano factor} $F \equiv S(0)/S_P$, which can vary between zero
and one.

Note that Nyquist and shot noises are in fact the two limiting
cases of the same phenomenon. One can express this noise in terms
of the transmission eigenvalues. There are other types of noise,
which are always or often present in nanostructures, and which are
not related to transmission properties. The most common example is
low-frequency noise, proportional to the square of the applied
voltage and inversely proportional to the frequency. The origin of
this noise is not universal and usually is attributed to slow
motion of impurities in the substrate. Such transport-unrelated
noises are not considered here.

Let us now mention basic properties of shot noise. More details
can be found in Ref. \cite{BB00} and references cited there.

\begin{itemize}

\item For basic types of nanostructures, the Fano factor assumes
universal values: $F=1$ for a tunnel barrier, $F = 1/2$ for a
symmetric double barrier, $F = 1/3$ for a diffusive wire, $F =
1/4$ for a symmetric chaotic cavity, $F=0$ for a ballistic
conductor (for instance, a quantum point contact in the plateau
regime). These results have been derived theoretically by various
means and confirmed experimentally.

\item These results are classical; quantum mechanics only enters
for calculation of transmission eigenvalues and in quantum (Fermi)
statistics of electrons. For this reason, many results can be
reproduced by purely classical methods, based on Boltzmann or rate
equations with Langevin random forces.

\item Notion of noise can be generalized to multi-terminal
conductors. Current correlations calculated at different terminals
are always negative. This follows from the fact that electrons
obey Fermi statistics.

\item The Fano factor is proportional to the electron charge. This
concept can be generalized to the situation when current is
carried by fermionic quasiparticles. For instance, transport
between a normal metal and a superconductor for voltages below the
superconducting gap is only possible by means of Andreev
reflection, and is associated with the charge transfer in quanta
of $2e$. This gives $F = 2$. Another example is transport in a
quantum Hall bar over a barrier, which is associated with the
charge $e/q$ for the filling factor $\nu = p/q$. The Fano factor
in this case becomes $F = 1/q$, which also has been measured
experimentally. Generally, shot noise can be used to determine the
quasiparticle charge.

\item Effect of interactions of shot noise can be very different.
Dephasing does not have any effect on noise, unless, of course,
one discusses a phase-sensitive effect like Aharonov-Bohm
oscillations. Electron-electron interactions result in heating,
increasing the Fano factor, for instance, to the value $F =
\sqrt{3}/4$ in diffusive wires instead of $F = 1/3$.
Electron-phonon interactions suppress shot noise down to zero,
since the energy is taken out of the system. This is why there is
no shot noise in macroscopic systems. All these considerations
assume that the ground state of a conductor is Fermi-liquid-like.
If interactions lead to a formation of a new state, the situation
can be very different.

\item Both shot and Nyquist noises are white --- frequency
independent in a wide interval of frequency. For non-interacting
electrons, the frequency dependence appears at the quantum scale
$\hbar \omega  \sim k_B T, e \vert V \vert$. For instance, at zero
temperature shot noise has the following form,
\begin{eqnarray}
S(\omega) = \frac{2_se^2}{\pi\hbar} \left\{ \begin{array}{lr}
\hbar \vert \omega \vert \sum_p T_p^2 + e \vert V \vert \sum_p T_p (1-
T_p), & \hbar \vert \omega \vert < e \vert V \vert; \\
\hbar \vert \omega \vert \sum_p T_p, & \hbar \vert \omega \vert > e
\vert V \vert.
\end{array} \right.
\end{eqnarray}
The part growing as $\vert \omega \vert$ is known as quantum
noise. Other energy scales come from electron-electron
interaction; in the most common case, the scale is just inverse
$RC$-time --- the time scale for classical charge relaxation.

\end{itemize}

Real life is fortunately more complicated than these simple rules,
and this is why current noise is still a subject of active
research. Below we consider a number of phenomena which go beyond
these rules and are currently in the focus of attention. We
specifically concentrate on four topics: non-symmetrized cumulants
and quantum noise; counting statistics; super-Poissonian noise;
and current noise and interferometry. For each of these subjects,
we outline the main ideas, describe the experiments, and provide
the full collection of references to the original papers.

Prior to that we would like to mention a number of experimental
developments of fundamental importance over the last five years.
They confirmed existed theoretical predictions, and generated a
subsequent stream of literature, but due to the space limitations
we can not discuss them in detail.

\begin{itemize}

\item Observation of shot noise suppression (cold and hot
electrons) in chaotic cavities \cite{cavity1}; crossover from
classical ($F=0$) to quantum ($F = 1/4$) shot noise in chaotic
cavities by tuning the dwell-time \cite{cavity2}.

\item Doubling of the Fano factor at the interface between a normal
metal and a superconductor \cite{Jehl}; also in the presence of
finite-frequency field (photon-assisted effect) \cite{Kozhevnikov}.

\item Very clean observations of giant shot noise in SNS junctions
due to multiple Andreev reflection (MAR)\cite{Cron}; crossover
from MAR regime to noise of the quasiparticle current
\cite{Lefloch1}; shot noise in the regime of coherent and
incoherent MAR in disordered SNS junctions \cite{Lefloch2}; MAR in
superconductor -- semiconductor -- superconductor junctions
\cite{Camino2}.

\item Noise in an array of quantum dots \cite{Nauen1}; noise in an
array of chaotic cavities formed by point contacts \cite{cavity3}.

\item Shot noise suppression in hopping conduction
\cite{Mendezhopping1,Camino1}.

\item Noise for photon-assisted tunneling \cite{Glattli1}.

\item Noise in quantum dots in the Coulomb blockade regime
\cite{Nauen2,Onac3}.

\end{itemize}

\section{Quantum noise}

In Eq. (\ref{noisedef}), we defined noise power as a symmetric
correlator. With this definition, $S(\omega)$ is always even in
frequency. Indeed, a classical detector does not know anything
about the order of the current operators and can not distinguish
between positive and negative frequencies: It only measures a
symmetric combination. Can we measure non-symmetrized noise,
\begin{equation} \label{nonsymmetr}
S_q (\omega) = 2 \int dt e^{-i\omega (t-t')} \left\langle \hat{I} (t)
\hat{I} (t') \right\rangle \ ?
\end{equation}

For such measurement, we obviously need a quantum detector. Let us
illustrate the basic notions with an example of a detector which
is a two-level system \cite{Aguado, Clerk}, with the states $\vert
a \rangle$ (energy $E_a$) and $\vert b \rangle$ (energy $E_b$).
Interaction between the detector and the system is supposed to be
weak and proportional to the current operator, $\hat{H} = \alpha
\vert b \rangle\langle a \vert \hat{I} (t) + h.c.$ The transition
rates between the two states of the detector follow from the Fermi
golden rule,
\begin{equation} \label{qnoiserates}
\Gamma_{a \to b} = \frac{\vert \alpha \vert^2}{2\hbar^2} S_q \left(
\frac{E_b - E_a}{\hbar} \right) \ .
\end{equation}
Thus, if one measures the transition rate {\em at zero frequency},
the result yields asymmetric noise correlator (\ref{nonsymmetr})
{\em at finite frequency}. For $E_b > E_a$, the detector absorbs
energy from the noise; otherwise, it emits energy. Thus, noise at
positive/negative frequency correspond to absorption/emission,
respectively. It is not symmetric, since transition rates are not
the same for emission and absorption. For instance, at equilibrium
these transition rates obey the detailed balance, $\Gamma_{a \to
b}p_a = \Gamma_{b \to a}p_b$, where $p_a$ and $p_b$ are the
occupation probabilities of the detector states, which obey
Boltzmann distribution. We obtain
\begin{displaymath}
S_q(\omega)/S_q(-\omega) = \exp(-\hbar\omega/k_BT) \ .
\end{displaymath}
At zero temperature, $S(\omega) = 0$ for positive frequencies:
There is no energy that detector can absorb from noise. For
non-equilibrium noise, at zero temperature absorption is only
possible if the energy provided by the external voltage is high
enough, $\hbar \omega < e \vert V \vert$. The result at zero
frequency expressed in terms of conductance $G$ and the Fano
factor $F$ of the nanostructure is
\begin{eqnarray} \label{asymneqnoise}
S_q(\omega) = 2G \left\{ \begin{array}{lr}
-2 \hbar \vert \omega, & \hbar \omega < -e \vert V \vert; \\
(e \vert V \vert - \hbar \omega) - (1-F)(e \vert V \vert + \hbar
\omega),\ \ \   & -e \vert V \vert < \hbar \omega < 0; \\
F(e\vert V \vert - \hbar \omega), & 0 < \hbar \omega < e \vert V
\vert; \\
0, & e \vert V \vert < \hbar \omega .
\end{array} \right.
\end{eqnarray}

First measurement of non-symmetrized noise using a quantum
detector was performed by Deblock {\em et al} \cite{OnacCPB} who
used a Josephson junction (JJ) as a detector. If JJ is biased at a
constant voltage, there is no dissipative (quasiparticle) current
at voltages lower than $2\Delta/e$, where $\Delta$ is the
superconducting gap. Such current could come from MAR, however, if
the transparency of the insulating layer between two
superconductors is low, these can be neglected. Thus, the
dissipative current is zero for $eV < 2\Delta$ and linearly grows
for higher voltages. However, if the junction is submitted to
external radiation, the situation is different: an electron can
absorb a phonon with the frequency $\omega$. Provided the voltage
is between $(2\Delta - \hbar\omega)/e$ and $2\Delta/e$, this
absorption will result in the quasiparticle dc current. The
amplitude of the current depends on the amplitude of the external
radiation. In particular, if the external radiation originates
from current noise produced by a nanostructure, the quasiparticle
current linearly depends on the {\em non-symmetrized} spectral
power of the current noise \cite{OnacCPB,Onac2},
\begin{displaymath}
I_{pat} (V) = \int_0^{\infty} \frac{d\omega}{2\pi} \left(
\frac{e}{\hbar\omega} \right)^2 S_{qV} (-\omega) I_{qp} \left( V +
\frac{\hbar\omega}{e} \right), \ \ \ eV < 2\Delta,
\end{displaymath}
where $I_{qp} (V)$ is the quasiparticle current without external
radiation, the photo-assisted current $I_{pat} (V)$ is the dc
current in the presence of external noise, and $S_{qV} (-\omega)$
is the non-symmetrized voltage noise, which is related to the
current noise via the impedance of the circuit. Sweeping the bias
voltage $V$ and measuring the current, one can restore the
frequency dependence of the non-symmetrized correlator. A great
advantage of such a detector is that the detector itself at $eV <
2\Delta$ is noiseless, since there is no quasiparticle current.

To demonstrate that the possibility of detection of non-symmetric
noise, Deblock {\em et al} \cite{OnacCPB} measured noise produced
by a Cooper pair box (also known as superconducting charge qubit).
This is a double junction superconducting structure with two close
levels corresponding to the states with $N$ and $N+1$ Cooper pairs
in the box, respectively. All other states of the system lie far
away from these two states and can be ignored. The splitting
$\epsilon$ between the states $N$ and $N+1$ can be tuned with the
gate voltage; the minimal value of this splitting is $\epsilon =
E_J$, with $E_J$ being the Josephson energy, achieved for $Q
\equiv C_g V_G = e$, where $C_g$ is the capacitance to the gate.
In an ideal system, the current noise is determined from Eq.
(\ref{qnoiserates}): only one transition is possible, with the
frequency $\epsilon/\hbar$, and thus the current noise has a delta
peak around this frequency. In reality, one has to take into
account that the levels are broadened by tunneling, and thus the
noise sharply peaks around the frequency $\omega_0 =
\sqrt{\epsilon^2 + \Gamma^2}/\hbar$, with $\Gamma$ being the
tunnel rate \cite{Plastina}. One speaks of the {\em quasiparticle
peak} in noise, with $Q < e$ and $Q > e$ corresponding to emission
and absorption, respectively. This is valid in the coherent regime
$E_J \gg \Gamma$; in the opposite incoherent regime, $E_J \ll
\Gamma$, one has a broad peak around zero frequency. The
experimental observation  of Ref. \cite{OnacCPB} was that noise on
the emission side of the quasiparticle peak is much stronger that
noise on the absorption side, thus confirming that the quantity
measured is the non-symmetrized current correlator.

A clear measurement of noise for a broad interval of frequencies
which could demonstrate the crossover from zero noise at the
absorption side to white shot noise at low frequencies and further
to quantum noise at the emission side is still not available in
the literature. However, there are further data demonstrating
detection of quantum noise. In Ref. \cite{Onac2}, one Josephson
junction was used as a noise source, and another one as a
detector. The source was biased at voltages above $2\Delta/e$ and
thus produced white shot noise from the quasiparticle current. The
detected noise was twice as low as the full expected shot noise of
the quasiparticle current, which shows that the non-symmetrized
correlator was measured: only $S_q(-\omega)$, but not the
contribution from $S_q(\omega)$.

Another detector used in the experiments is a quantum dot
\cite{Hartmann} at low bias. Without external noise, current only
flows when an electron level lies in the window between the
chemical potentials of the reservoirs. As the function of the gate
voltage, current has a peak. With the external noise, tunneling
via excited states becomes possible, and additional peaks in the
current appear. In the experiments, the magnitude of these
additional peaks at the emission side was clearly stronger than
the one at the emission side.

\section{Counting statistics}

Shot noise originates from random nature of electron transfers.
One can, at least in principle, count these transfers in real
time, and from the results of the measurements deduce average
current and current noise. We have seen that shot noise contains
some information about scattering properties of the nanostructure
which can not be obtained from the conductance. Higher current
moments can also be deduced from the same measurement and  may
contain even more information. They are described with the notion
of {\em full counting statistics} (FCS).

Let us proceed with a bit of the probability theory. Suppose
we make a measurement counting some random events --- for instance,
electron transfers through a barrier --- during a certain
time interval $\Delta t$. The number of events $N$ measured
during the time interval is a {\em random number}, characterized by
the probability $P_N$ that precisely $N$ events will be observed in a
measurement. If one repeats identical measurements $M_{tot}$ times and
counts the number of measurements $M_{N}$ that give the count $N$, the
ratio $M_{N}/M_{tot}$ gives the probability $P_{N}$ in the limit
$M_{N} \gg 1$. This probability distribution is normalized, $\sum_N
P_N = 1$. Once we know it, we can estimate the average of any function
$f_N$,
\begin{displaymath}
\langle f \rangle = \sum_N f_N P_N.
\end{displaymath}
The description of the statistics with the distribution function $P_N$
is not always the most convenient one. The problem is that if we
measure first during the time interval $\Delta t_1$ (distribution
function $P_1$) and then during $\Delta t_2$ ($P_2$), the
distribution function for the total interval $\Delta t = \Delta t_1 +
\Delta t_2$ is a convolution (provided the two intervals are independent),
\begin{displaymath}
P^{tot}_N = \sum_{M=0}^{N} P_{1,M} P_{2, N-M}.
\end{displaymath}
Most conveniently, this is expressed in terms of {\em characteristic
function} of a probability distribution,
\begin{displaymath}
\Lambda (\chi) = \left\langle e^{i\chi N} \right\rangle = \sum_N
P_N e^{i\chi N}.
\end{displaymath}
For independent events, the characteristic function of the total
distribution is just a product of characteristic functions of each
type of events, $\Lambda^{tot}(\chi) = \Lambda_1(\chi)
\Lambda_2(\chi)$. The function $\ln \Lambda(\chi)$ is thus
proportional to the duration of the measurement $\Delta t$.
Differentiating this function $k$ times with
respect to $i\chi$ and setting subsequently $\chi = 0$, we generate
the $k$th cumulant of the distribution. Thus, the first derivative
produces the average $N$, and the second derivative reproduces the
variance.

For electron transfers in nanostructures, it is customary to consider
statistics of charge $Q = eN$ transmitted from the left to the right
during the time interval $\Delta t$. We assume that this measurement
time is long enough, so that $Q \gg e$ and the laws of statistics
apply. On average, $\langle Q \rangle = \langle I \rangle \Delta
t$. Second cumulant gives the shot noise at zero frequency,
\begin{displaymath}
\langle\langle Q^2 \rangle\rangle \equiv \langle Q^2 \rangle - \langle
Q \rangle^2 = \Delta t S(0)/2.
\end{displaymath}

The characteristic function of the transmitted charge can be expressed
in terms of transmission eigenvalues of the nanostructure
\cite{LevLes1,Yakovets,LevLes2},
\begin{eqnarray} \label{Levitov}
\ln \Lambda(\chi) & = & 2_s\Delta t \int \frac{dE}{2\pi\hbar} \sum_p \ln
\left\{ 1 + T_p \left( e^{i\chi} - 1 \right) f_L (E) \left[ 1 - f_R
(E) \right] \right. \nonumber \\
& + & \left. T_p \left( e^{-i\chi} - 1 \right) f_R (E) \left[ 1 -
f_L (E) \right] \right\}.
\end{eqnarray}

The logarithm of characteristic function is a sum over transport
channels, this suggests that electron transfers in different channels
and over different energy intervals are independent. Differentiating
this expression once and twice over the counting field $\chi$, we
reproduce Landauer formula and the expression for the current
noise. At zero temperature, Eq. (\ref{Levitov}) becomes
\begin{equation} \label{Levitov0}
\ln \Lambda(\chi) = \pm \frac{2_seV\Delta t}{2\pi\hbar} \sum_p \ln
\left[ 1 + T_p \left( e^{\pm i\chi} - 1 \right) \right],
\end{equation}
where the upper and lower signs refer to the case of positive and
negative voltages, respectively. Let us for simplicity consider $V
> 0$. We define the number of attempts
$N_{at} = 2_s \Delta t eV/(2\pi\hbar)$ and assume it to be
integer. Eq. (\ref{Levitov0}) for one transport channel
corresponds to the binomial distribution,
\begin{equation} \label{binomial}
P^{(p)}_N = \left( \begin{array}{c} N_{at} \\ N \end{array} \right)
T_p^N (1-T_p)^{N_{at}-N}.
\end{equation}
This is just the probability that out of $N_{at}$ electrons arriving
to the barrier $N$ pass through, and others, $N_{at} - N$, are
reflected back. For more than one channel, the binomial distribution
does not hold any more: One obtains a convolution of binomial
distributions corresponding to each channel. If all
transmission eigenvalues are small, Eq. (\ref{binomial}) yields the
Poisson distribution, corresponding to the notion of independent
electron transfers.

One can now averaging Eq. (\ref{Levitov0}) over various distributions
of transmission eigenvalues to produce the full counting
statistics. In this way, FCS for a double barrier \cite{deJong},
diffusive wires \cite{Yakovets,NazarovAnn}, and chaotic cavities
\cite{Schomerus} was produced.

The concept of FCS can be generalized to different situations,
where Eq. (\ref{Levitov}) does not apply any more. By now, full
counting statistics is a field by itself. Various methods have
been developed to calculate higher cumulants, which include
semi-classical approach based on the cascade of Langevin
equations, both for diffusive wires \cite{NagaevCumul} and chaotic
cavities \cite{Nagaev1}, semi-classical circuit theory based on
Keldysh Green's function technique
\cite{BelzigNazarov1,Bagrets1,Bagrets2}, semi-classical stochastic
path integral method \cite{Pilgramstoch}, Keldysh sigma-model
\cite{Gutman1,Gutman2}, direct numerical simulation \cite{Pala},
and analytical\cite{Roche1} and numerical\cite{Roche2} treatment
of exclusion models. Results on FCS for normal-superconductor
interfaces in various situations
\cite{Muz,BelzigNazarov1,Samuelsson3}, for charge transfer between
superconductors, both for applied constant phase
\cite{NazarovBelzig2} or applied constant voltage
\cite{Belzig4,Belzig5,Johansson}, for quantum to classical
crossover in chaotic cavities \cite{cavity5,cavity6}, for FCS in double
quantum dots \cite{Wacker2}, in
multi-terminal circuits, including superconducting elements
\cite{Bagrets1,Borlin}, quantum dots in the Coulomb blockade
regime \cite{Bagrets2,Bagrets4}, interacting diffusive conductors
\cite{Gutman1,Pilgram3,Bagrets3,Gutman2}, for frequency dependence
of the higher cumulants \cite{Galaktionov,Nagaev3,Nagaev4}, and
for the time-dependent current \cite{Ivanov1,Ivanov2} are
available. There is one issue we would like to mention here. For
FCS in the charge transfer between two superconductors with the
fixed phase difference, the probabilities $P_N$ can sometimes
assume negative values \cite{NazarovBelzig2}. The reason is that
phase and charge are canonically conjugated variables and can not
be measured simultaneously. No such problem exists for
voltage-biased junctions. Thus, in this case the quantities $P_N$
can not be interpreted as probabilities. However, Ref.
\cite{NazarovBelzig2} suggested a scheme that makes a measurement
of $P_N$ possible, even if they are negative. The set of the
quantities $P_N$ contains full information about the charge
transfer in the system even in this situation.

Let us now specifically consider the third cumulant of transmitted
charge. It has a very important property:  At equilibrium
($V=0$) the characteristic function of Eq. (\ref{Levitov}) becomes
even in $\chi$, and therefore all the odd cumulants disappear. There
is no ``Nyquist third cumulant''. For this reason, one does not have
to measure at very low temperatures to extract the information of
transmission eigenvalues. On the other hand, noise measurement is
already more complicated than the average current measurement since it
requires to collect more measurement results to achieve decent
accuracy. The direct measurement of the third cumulant is even more
challenging.

From Eq. (\ref{Levitov0}) we get the third cumulant in the ``shot
noise'' regime $eV \gg k_BT$,
\begin{equation} \label{av3rdcum}
\langle\langle Q^3 \rangle\rangle = e^3 \frac{\partial^3}{\partial
(i\chi)^3} \ln \Lambda(\chi) = e^2 V G_Q \Delta t \sum_p T_p (1 -
T_p) (1 - 2T_p).
\end{equation}
For a tunnel barrier $T_p \ll 1$ we get $\langle\langle Q^3
\rangle\rangle = e^2 \langle I \rangle \Delta t$, which can be
derived directly from the Poisson distribution. In a diffusive
wire, the averaging over the distribution function of transmission
eigenvalues yields $\langle\langle Q^3 \rangle\rangle = e^2
\langle I \rangle \Delta t/15$. The third cumulant can be either
positive or negative, the open channels with $T_p > 1/2$ favour
negative sign. One can also derive the full voltage dependence,
including the regime $eV \ll k_BT$, from Eq. (\ref{Levitov}).

\begin{figure}
\centering
\includegraphics[width=7cm]{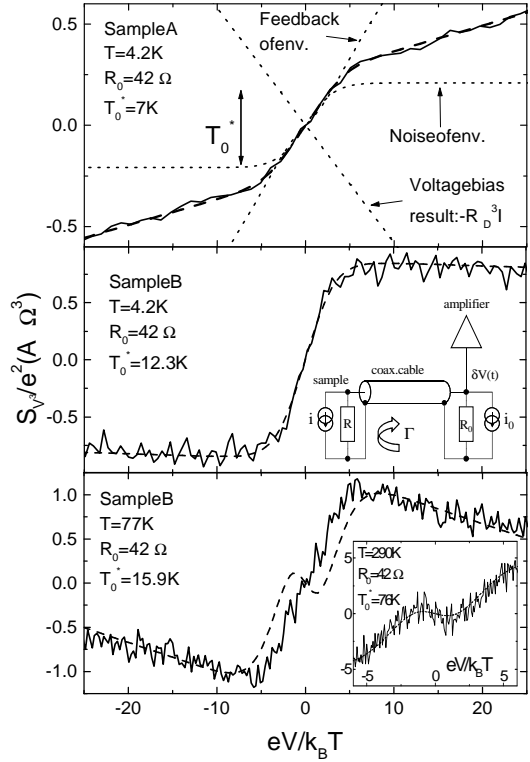}
\caption{Measurements of the third cumulant \protect\cite{Reulet}.
The top panel represents comparison of the results with the
theory. The agreement is only achieved if the enviromental
fluctuations are taken into account, see Eq.
(\protect\ref{thirdcumul}). Copyright (2003) by American Physical
Society.} 
\label{fig:Reulet} \end{figure}

The first measurement of a higher cumulant of current noise was
performed by Reulet, Senzier, and Prober \cite{Reulet}, who
studied the third cumulant of the voltage drop $S_V^3$ across a
tunnel junction biased by a constant current. The result is
plotted in Figure ~\ref{fig:Reulet} against the average voltage
drop over the junction (the resistance of the circuit $R_D$ is
found independently from the noise measurements). One now has to
compare experimental measurements with the theoretical prediction
(\ref{av3rdcum}). Naively, third cumulant of the voltage is
obtained from the third cumulant of the current $S_I^3$ by
multiplication with $R_D^3$. This operation produces a dashed line
shown in Figure~\ref{fig:Reulet} as ``voltage bias result''
--- in strong contradiction with experimental data. The reason for this
disagreement was discovered in Refs. \cite{Kindermannenv}. It is
known that, similarly to the average voltage $V = R_DI$, voltage
fluctuations can be obtained from the current fluctuations by
multiplication with $R_D^2$, $S_V = R_D^2 S_I$. However, for third
and higher cumulants the situation is more complicated, since
voltage fluctuations generated at the sample can perturb other
parts of the electric circuit and generate there current
fluctuations, which affect current fluctuations at the sample.
This strongly modifies the relation between current and voltage
third cumulants. In the simplest situation, when the environment
itself is noiseless (``non-invasive measurement'') one has
\begin{equation} \label{thirdcumul}
S_{V^3} = -R_D^3 S_{I^3} + 3R_D^4 S_I \frac{dS_I}{dV}.
\end{equation}
The difference between higher order cumulants is even more
dramatic. To characterize voltage fluctuations, one defines the
phase $\phi = \int_0^{\Delta t} \delta V (t) dt$ and studies full
counting statistics for the phase. It turns out that the phase has
Pascal distribution rather than binomial distribution
(\ref{binomial}) which one finds for the transmitted charge; in
particular, for the tunnel junction, the distribution is chi
square rather than Poisson one \cite{chisquare,Kindermann2}.

The problem with the experiment \cite{Reulet} is that the measured
third cumulant was dominated by the shot noise contribution
(second term in Eq. (\ref{thirdcumul})), and the contribution
$S_{I^3}$ only became important at high temperatures, when the
voltage dependence of the noise is weak. To avoid this, several
groups performed measurements of the transmitted charge in real
time. Following an earlier theoretical suggestion of Ref.
\cite{Reznikov04}, Bomze {\em et al} \cite{Bomze} measured the
third current cumulant of a tunnel junction by amplifying and
analyzing in real time voltage fluctuations on a detector --- a
resistor with the conductance much higher than that of the sample.
The measurements were performed at $4K$ and demonstrated the
dependence $S_{I^3} = e^2 \langle I \rangle$, expected from the
Poisson distribution.

Real-time measurements are easier in the Coulomb blockade regime in
quantum dots, since the electron spends a relatively long time in the
system, electrons enter the dot one by one, and individual tunnel
processes are easier to resolve. The earlier measurements \cite{Rimberg}
used a single-electron transistor electrostatically coupled to the
quantum dot as a detector, and observed real-time detection of single
electron tunneling; however, the measurement precision was not enough
to extract full counting statistics. Subsequently, more precise
real-time detector measurements were performed in quantum dots
\cite{Fujisawa1} and in superconducting junction arrays
\cite{Bylander}, still without extracting the full counting
statistics. Then Gustavsson {\em et al} used charge detection with a
quantum point contact electrostatically coupled to the quantum dot:
When an electron enters the dot, it increases the height of the
potential barrier in the quantum point contact. If the detector is
tuned close to the step between the plateaus, this increase would
block electron passage through the junction. Thus, measurement of the
current through the detector gives the real-time information on the
occupation of the quantum dot. They plotted the hystogram of the
transmitted charge during the measurement time $\Delta t$ (half a
second in their experiment) and analyzed the FCS. Very recently
Fujisawa {\em et al} \cite{Fujisawa2} reported measurements of FCS in
a double quantum dot, also with a quantum point contact as a
detector. They collected enough statistics to restore the rates
for all possible tunnel processes, checked the detailed balance
relation between the rates, produced the occupation probabilities by
solving the master equations, and compared the results with the
observed FCS.

All these recent experimental advances concentrate on the
situations when the FCS from the point of view of the theory is
trivial --- Poisson distribution in tunnel junctions, and only two
possible charge states in quantum dots --- and so far serve rather
to demonstrate that FCS can be measured. Measurements of less
trivial effects, for instance,  of tails of the distribution
function of the transmitted charge, are still to come.

\section{Super-Poissonian noise}

It follows from Eq. (\ref{shotres}) that shot noise in the system
of non-interacting electrons is always sub-Poissonian: the Fano
factor $F$ is less than one. This means that every time
super-Poissonian noise is measured the reason must be looked for
in interactions (typically electron-electron interactions).
However, this statement is too general, and, as many too general
statements, useless. Let us look in more details at different
situations which can produce super-Poissonian noise.

As we mentioned, shot noise measures charge quantum transferred
across the nanostructure. An example we already mentioned is an
interface between a normal metal and a superconductor. For
voltages below the superconducting gap transport is only possible
by Andreev reflection, and the corresponding charge quantum is
$2e$. The Fano factor for such system can be up to $F = 2$.
Another example is transport in SNS systems, which proceeds via
multiple Andreev reflections. Such process is associated with
transfer of $\Delta/eV$ charge quanta, which provides
super-Poissonian Fano factors.

\begin{figure}
\centering
\includegraphics[width=7cm]{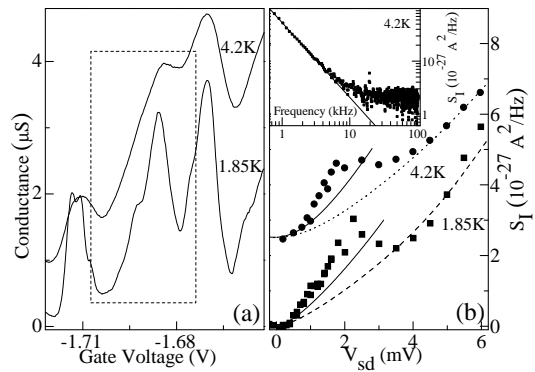}
\caption{Tunneling via localized states \protect\cite{Safonov}. Noise
is shown at the right panel; solid line represent Poisson
value. Copyright (2003) by American Physical 
Society.}
\label{fig:Safonov}
\end{figure}

Let us give another example illustrating the same mechanism.
Safonov {\em et al} \cite{Safonov} studied noise in tunneling via
localized states. They discovered that the Fano factor strongly
depends on the gate voltage, sometimes achieving values above one
(Figure \ref{fig:Safonov}). To explain these results, Ref.
\cite{Safonov} suggested the following model. Imagine the
transport occurs through two localized states (``impurities'') in
parallel: $R$ and $M$. $M$ is coupled to the leads much weaker
than $R$, so that its contribution to the current and noise is
negligible. If $R$ and $M$ were independent, the Fano factor would
vary between $1/2$ and $1$, depending on the asymmetry of the two
barriers separating the impurity from the reservoirs. However,
things change if the two impurities are coupled electrostatically.
Due to this coupling, the occupation of $M$ affects the position
of electron levels in $R$ and can, for instance, shift a level off
the resonance, blocking the transport through $R$. In this case,
if $M$ is occupied, current through $R$ is blocked, and if $M$ is
empty, current through $M$ proceeds in an ordinary way. Thus, if
tunnel rates for $M$ and $R$ are of the order of $\Gamma_R$ and
$\Gamma_M$, respectively, transport through the system proceeds in
bunches of $\Gamma_R/\Gamma_M \gg 1$ electrons. Thus, the Fano
factor can achieve large super-Poisson values. Since such a
two-impurity configuration is not expected to be typical, after
impurity averaging, shot noise is considerably reduced, which
explains Fano factors slightly above one observed in the
experiment.

This example illustrates a general mechanism of super-Poissonian
noise. If transport proceeds via two or more electrostatically
coupled states, so that occupation of one of the states ($M$) may
block the transport through the other one(s), $R$, charge is
transported in quanta of $e\Gamma_R/\Gamma_M$. Super-Poissonian
noise appears provided this number is greater than one, that is
the states are coupled differently to the leads. This situation
may occur in quantum dots in various regimes under the Coulomb
blockade condition: One needs that the charging energy is greater
than the separation between the energy levels relevant for
transport ({\em dynamical channel blockade}). Theoretical
predictions of super-Poissonian noise exist for sequential
tunneling regime in quantum dots with ferromagnetic leads
\cite{Bulka,Cottet2,Braun} (if both leads are partially polarized
say spin-up, then spin-up electrons tend to tunnel in bunches, and
spin-down electrons block the current for a long time), in a
magnetic field \cite{Cottet1}, general dynamical channel blockade
for sequential tunneling
\cite{Wacker,Cottet3,Thielmann,Belzigseqtun}, in double quantum
dots \cite{Djuric}, and in quantum dots where the level coupling
is mediated by non-equilibrium plasmons in the leads
\cite{Kinaret}. Ref. \cite{Burkard} predicted super-Poissonian
noise in the inelastic cotunneling regime.

Recently super-Poissonian noise was experimentally observed by Onac
{\em et al} \cite{Onac3} in a carbon nanotube quantum dot (carbon
nanotube crossed by two barriers). They measured noise across the
Coulomb diamond and found Fano factors up to $F=3$. Super-Poissonian
noise was observed inside the diamonds and is therefore associated
with inelastic cotunneling.

Another source of super-Poissonian noise is bistability. An
example was provided by Refs. \cite{Iannaccone,Mendez1}, that
studied current through quantum wells in the resonant tunneling
regime. A similar behavior was observed in tunneling through a
zero-dimensional state \cite{Nauen3}. Due to interactions, in a
certain interval of voltages these wells become bistable: One
state with zero current and one state with finite current. For
lower voltages, zero-current branch becomes unstable and ceases to
exist; for higher voltages, finite-current branch does not exist.
Noise in such system comes from the two sources: (i) ``shot''
noise --- small current fluctuations around each of the branches;
(ii) random jumps --- random telegraph noise
--- between the branches. Close to the instability point, when the
finite-current branch disappears, current fluctuations around this
branch diverge and exceed the Poisson value \cite{BB99,BB00}. Full
analysis of the noise also includes large fluctuations, resulting
in jumps between the branches \cite{Tretiakov1,Tretiakov2}. Ref.
\cite{Jordan1} treats current statistics in a generic bistable
system. Experimentally, a link between super-Poissonian noise and
bistability is not established convincingly: For instance, a
superlattice tunnel diode is a bistable system, but experiments
did not discover any super-Poissonian noise enhancement
\cite{Mendez2}.

Finally, we discuss yet another situation, when there are other
degrees of freedom in the system which affect transport
properties, in particular, noise. Actually, this situation is
rather common: In any mesoscopic and macroscopic system, electrons
interact with phonons. If this interaction is effective enough ---
the characteristic length of electron-phonon relaxation $L_{ph}$
is shorter than the size of the system --- electrons relax to the
equilibrium distribution, and the noise they produce is Nyquist
noise --- shot noise disappears. If we want to have anything
non-trivial, we need to consider non-equilibrium phonons. Such
opportunity was recently provided by a new class of devices ---
nanoelectromechanical systems (NEMS), which couple electron motion
to mechanical degrees of freedom. Currently, many species of NEMS
were made and investigated, including shuttles --- single-electron
tunneling devices with movable central island, double-clamped
suspended beams in the Coulomb blockade regime, or single-clamped
suspended cantilevers. First, in NEMS one can discuss not only
charge noise, but also momentum noise --- random fluctuations of
momentum transferred from electrons to the crystalline lattice
\cite{Shytov,KindermannNems,Tajic}. Second, ordinary current noise
is strongly modified by mechanical degrees of freedom, both in the
shuttling regime \cite{Jauho1,Nord,Pistolesi,Jauho2} and for
single-electron tunneling
\cite{Zhu,Armour1,Bruder1,Koch,Yu,Armour2}; in particular, in both
situations noise can achieve super-Poisson values. It is not our
intention to give here a comprehensive review of noise in NEMS,
and we restrict ourself to just one particular situation: a single
electron tunneling device weakly coupled to a single-mode
underdamped harmonic mechanical oscillator\cite{Usmani1,Usmani2}.

Qualitatively, the situation is as follows. Imagine that bias and
gate voltages are tuned just outside a Coulomb blockade diamond,
so that only two charge states are important for electrons: say
$n=0$ and $n=1$. Non-zero average current means that the number of
electrons in the single-electron tunneling device fluctuates
randomly between zero and one. The coupling between electrons and
vibrations of the oscillator is provided by a force $F_n$ which
depends on the charge state and acts on the oscillator. In the
regime we discuss this force is a random function that can assume
two values, $F_0$ and $F_1$. The force generates mechanical
oscillations, which in the underdamped case have a large amplitude
and a frequency close to the eigenmode $\omega_0$ of the
oscillator. The vibration produces feedback on the current since
the tunnel rates depend on the position of the oscillator, due to
the position dependence of the energy differences available for
tunneling. It turns out that the quality factor of the oscillator
$Q$ is renormalized due to electron tunneling, but even after the
renormalization one still is in the underdamped regime.

Depending on the voltages, one can identify four types of the
behavior of the system, which are best described in terms of the
probability $P(A)$ to have certain amplitude $A$ of the mechanical
oscillator. First, $P(A)$ can be sharply peaked around zero
(meaning only very small amplitudes have significant probability)
or around certain finite value. In both cases, noise can be
estimated as follows. From dimensional analysis, one obtains $S(0)
\sim I^2 \tau$, where $\tau$ has the dimensions of time. In the
ordinary situation, $\tau$ is of order of the inverse tunnel rate
$\Gamma^{-1}$ (the only energy scale in the problem), and one
restores the Poisson value of the shot noise. In our case, there
is a longer time scale --- the decay time $Q/\omega_0 \gg
\Gamma^{-1}$; thus, we have $S(0) \sim eI \Gamma Q/\omega_0$, and
noise considerably exceeds the Poisson value. In two further
cases, the distribution function $P(A)$ has two peaks: either one
at $A = 0$ and another one at a finite value of $A$, or both peaks
at finite values of $A$. This means that only two values of the
amplitude are possible. In both cases, on top of super-Poissonian
noise for each peak, we have additional enhancement of noise due
to random jumps between the states with different values of the
amplitude.

\section{Interference effects}

Interference effects are at the core of quantum mechanics: They
provide information on the phase of the wave function probing its
wave nature. These experiments are difficult to perform with
electrons in solids, however, we witness steady progress over the
last decade, with a number of proposals and successful
realisations constantly increasing. We will only discuss here the
types of interferometers for which studies of noise are available.

Qualitatively the simplest species is an Aharonov-Bohm (AB) ring
--- a two-terminal structure where the transmission probability is
a periodic function of the magnetic flux penetrating the ring,
with the period of the flux quantum. Conductance and shot noise
retain this periodic dependence. However, this system has just too
few handles, the amplitude of AB oscillations depends essentially
on the dynamical phases acquired by electrons moving along the
arms of the ring, and the AB vanishes in a ring already with
several transport channels.

\begin{figure}
\centering
\includegraphics[width=7cm]{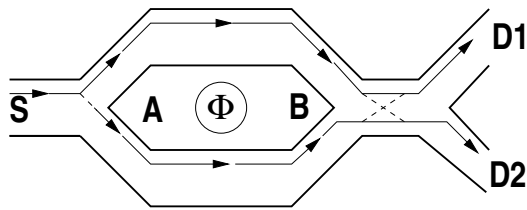}
\caption{Mach-Zehnder interferometer with edge states. Edge states
are shown by solid lines with arrows; additional tunnel directions
in the beam splitters with dashed lines.} \label{fig:MZI}
\end{figure}

Recently, Ji {\em et al} \cite{Heiblum} performed an experiment
with the electronic Mach-Zehnder interferometer (MZI), an analog
of the corresponding optical device. The electronic MZI with edge
states in the integer quantum Hall regime is shown in Figure
~\ref{fig:MZI}. In the simplest version, it has one source and two
detectors (the voltage $V$ is applied to the source relative to
both detectors), the beam splitter A and the beam splitted B, both
realized as quantum point contacts. The transport proceeds via the
edge states; we assume that in the contact A there is no
reflection, so that the edge state proceeds either to the upper or
to the lower arm. In the beam splitter B, there is also no
reflection, and an electron from either arm can proceed to one of
the detectors, $1$ or $2$. An Aharonov-Bohm flux penetrates the
ring. Both conductance between the source and any of the
detectors, and current correlations between any reservoirs are
flux-dependent. In the experiment, transmission through one of the
point contacts could be changed, which creates an additional
handle.

Surprisingly, the visibility --- the ratio of the phase-dependent
and phase-independent parts of the conductance --- observed in the
experiment, was lower than expected. This suppression of the
visibility can be attributed to the loss of phase coherence, which
partially destroys the interference. There could be two reasons
for this loss of coherence: phase averaging (for instance, due to
the energy dependence of the phase) and dephasing by environment,
in particular, by electron-electron interactions. Conductance is
affected by both mechanisms in the same way: It is proportional to
the transmission probability from the source to the corresponding
drain, averaged over the phase. The details depend on the type of
the dephasing \cite{Seelig1,Seelig2}. Thus, conductance generally
can not distinguish between these two reasons of phase coherence
loss, if they have the same dependence of the dephasing time
(usually it is proportional to the temperature). It turns out that
the situation is different for noise \cite{Marquardt1,Marquardt2},
as well as for higher cumulants of the transmitted charge
\cite{Forster}. One has to compare the applied voltage $eV$ with
the inverse characteristic lifetime for the phase correlations
induced by the environment $\hbar/\tau_c$. In this way, one
identifies ``fast'' $eV \tau_c \ll \hbar$ and ``slow'' $eV \tau_c
\gg \hbar$ environments. For slow environment, shot noise is
merely obtained by averaging of the usual expression, $T (1 - T)$,
over the phase. In this case, it does not provide any new
information on the environment as compared to the conductance. For
fast environments, the behavior of noise is generally different,
and thus may provide an information on the source of loss of the
phase coherence.

In Refs. \cite{DMZ1,DMZ2}, a more complicated two-particle
interferometer was proposed. It consists of four sources and four
detectors, separated by four quantum point contacts. Transport is
again only possible via the edge states in the integer quantum
Hall regime. The setup is designed in such a way that an electron
can be transmitted from any source to any detector via {\em only
one} trajectory. Then, average current is not sensitive to
interference, does not depend on the Aharonov-Bohm phase, and is
determined by transmission probabilities of the contacts. In
contrast, if we consider current cross-correlations at different
detectors, the interference contribution originates from the
interference of {\em different} trajectories. In this 
geometry, these pairs of trajectories are taken in
such a way that {\em together} they enclose a loop and are thus
sensitive to the Aharonov-Bohm flux. This is probably up to now
the nicest illustration of the two-particle nature of current
noise in solid state systems.

We last mention an Andreev interferometer --- a normal metal
connected by two arms with the same superconducting reservoir.
Transport properties of this system are sensitive to magnetic flux
enclosed by the arms. Reulet {\em et al} \cite{AndInterf1}
investigated current noise in this system experimentally and
theoretically, and found periodic dependence on the applied flux.
Full counting statistics in an Andreev interferometer was studied
in Ref. \cite{Bezugly}.

\section{Acknowledgements}

The author very much appreciates collaboration with his friends
and colleagues on various issues related to current noise, in
chronological order: Stijn van Langen, Eugene Sukhorukov, Henning
Schomerus, Carlo Beenakker, Gabriele Campagnano, Oleg Jouravlev,
Yuli Nazarov, Omar Usmani, Thomas Ludwig, Alexander Mirlin, and
Yuval Gefen. He is especially grateful to Markus B\"uttiker, who
introduced him to the field of shot noise.



\printindex
\end{document}